\documentclass[twocolumn, amsmath, amssymb]{revtex4}

\usepackage{siunitx}
\usepackage[dvipdfmx]{graphicx}

\DeclareMathOperator{\erf}{erf}
\DeclareMathOperator{\erfc}{erfc}

\newcommand{\LN}{\ln\mathcal{N}}

\begin{document}
\title{Analysis of a Stochastic Model for Bacterial Growth and the Lognormality of the Cell-Size Distribution}
\author{Ken Yamamoto\thanks{yamamoto@phys.chuo-u.ac.jp} and Jun-ichi Wakita}
\affiliation{Department of Physics, Faculty of Science and Engineering, Chuo University, Kasuga, Bunkyo, Tokyo 112--8551, Japan}
\begin{abstract}
This paper theoretically analyzes a phenomenological stochastic model for bacterial growth.
This model comprises cell division and the linear growth of cells, where growth rates and cell cycles are drawn from lognormal distributions.
We find that the cell size is expressed as a sum of independent lognormal variables.
We show numerically that the quality of the lognormal approximation greatly depends on the distributions of the growth rate and cell cycle.
Furthermore, we show that actual parameters of the growth rate and cell cycle take values that give a good lognormal approximation; thus, the experimental cell-size distribution is in good agreement with a lognormal distribution.
\end{abstract}

\maketitle

\section{Introduction}
\subsection{General overview}
In the study of complex phenomena, we can extract statistically useful information from the size distribution of observed elements.
Along with the power-law distribution\cite{Newman, Buchanan}, which is closely associated with critical phenomena in statistical physics, the lognormal distribution~\cite{Crow} is found in a wide range of complex systems.~\cite{Kobayashi, Limpert}
It appears in natural phenomena such as fragment and particle sizes\cite{Katsuragi, Bittelli}, the fluctuation of X-ray bursts\cite{Uttley}, and genetic expression in bacteria\cite{Furusawa},
and in social phenomena such as population\cite{Sasaki}, citations of scientific papers\cite{Redner}, and proportional elections\cite{Fortunato}.
The commonness of the lognormal behavior is helpful for studying various complex systems from a unified viewpoint based on statistical physics.

A random variable $X$ is said to follow a lognormal distribution $\LN(\mu, \sigma^2)$ if its logarithm $\ln X$ is normally distributed with mean $\mu$ and variance $\sigma^2$.
The probability density of $\LN(\mu, \sigma^2)$ is
\[
f_{\mu,\sigma}(x)=\frac{1}{\sqrt{2\pi\sigma^2}\, x}\exp\left(-\frac{[\ln x-\mu]^2}{2\sigma^2}\right),
\]
and the (upper) cumulative distribution function is
\begin{align}
F_{\mu,\sigma}(x)&=\int_x^\infty f_{\mu,\sigma}(y)dy\notag\\
&=\frac{1}{2}\left[1-\erf\left(\frac{\ln x-\mu}{\sqrt{2}\, \sigma}\right)\right]
=\frac{1}{2}\erfc\left(\frac{\ln x-\mu}{\sqrt{2}\, \sigma}\right),
\label{eq1}
\end{align}
where $\erf(x)=(2/\sqrt{\pi})\int_0^x \exp(-y^2)dy$ is the Gauss error function
and $\erfc(x)=1-\erf(x)$ is the complementary error function\cite{Abramowitz}.
The lognormal distribution is a heavy-tailed distribution; that is, the tail of $F_{\mu,\sigma}(x)$ decays slower than any exponential function.
Thus, a phenomenon obeying a lognormal distribution easily produces values much larger than the mean.
The mean of the distribution $\LN(\mu, \sigma^2)$ is $\exp(\mu+\sigma^2/2)$ and the median is $\exp(\mu)$.
The mean is always larger than the median, and this is a consequence of the heavy-tailed property of the lognormal distribution.\cite{Sornette}

The lognormal distribution is typically generated by a multiplicative stochastic process.\cite{Mitzenmacher}
Consider a stochastic process $X_1, X_2,\dots$ given by
\[
X_{n+1}=M_nX_n,
\]
where $M_n$ is a positive random variable.
If the growth rates $M_1,M_2,\dots$ are independently and identically distributed, and $E[\ln M_n]=\mu$ and $V[\ln M_n]=\sigma^2$ are finite,
the central limit theorem implies that the distribution of $(\ln X_n-n\mu)/(\sqrt{n}\sigma)$ converges to the standard normal distribution (with zero mean and unit variance).
Roughly speaking, the distribution of $X_n$ for sufficiently large $n$ is reasonably approximated by the lognormal distribution $\LN(n\mu, n\sigma^2)$.
The lognormal parameters $n\mu$ and $n\sigma^2$ diverge as $n\to\infty$; thus, $X_n$ itself does not have a stationary distribution exactly.
Many systems have multiplicativity in some sense; thus, the lognormal distribution is a natural distribution for their statistical properties\cite{Kobayashi}.

It should be checked carefully whether a probability distribution obtained from actual data is truly lognormal or not, even if the lognormal fitting is visually successful.
Many researchers have pointed out that a lognormal distribution can be confused with other probability distributions such as power-law\cite{Malevergne}, normal\cite{Kuninaka}, and stretched exponential\cite{Stumpf} distributions.
Furthermore, the multiplicative nature does not necessarily lead to a lognormal distribution.
In fact, lognormal behavior can easily change into other distributions if additional rules are assigned to the multiplicative stochastic process.
For instance, multiplicative processes with additive noise\cite{Takayasu}, reset events\cite{Manrubia}, random stopping\cite{Yamamoto2012} and successive sum\cite{Yamamoto2014} produce power-law distributions.
It is difficult to examine whether an experimental data set follows a genuine lognormal distribution or a lognormal-like distribution.

In this paper, we perform a theoretical analysis of a stochastic model of bacterial cell size.
Wakita \textit{et al.}\cite{Wakita} reported that the cell-size distribution of the bacterial species \textit{Bacillus}~(\textit{B.})~\textit{subtilis} is well described by a lognormal distribution.
They also proposed a phenomenological stochastic model for the growth of bacterial cells and confirmed that the numerical result quantitatively reproduces the actual cell-size distribution.
In contrast to these results, we show in this paper that the cell-size distribution generated by this model is not a genuine lognormal distribution in reality.
In order to elucidate the situation and the problem, we give an outline of the experiment and model in the next subsection.

\subsection{Experiment and modeling of bacterial cell size}
As a background of the present paper, we concisely review the above-mentioned experimental study~\cite{Wakita} regarding the cell-size distribution of \textit{B.~subtilis}, which is a rod-shaped bacteria of 0.5--\SI{1.0}{\micro\metre} in diameter and 2--\SI{5}{\micro\metre} in length.
The morphology of a \textit{B.~subtilis} colony spreading on an agar surface changes with the nutrient and agar concentrations and is classified into five types:
diffusion-limited aggregation-like, Eden-like, concentric ring-like, homogeneously spreading disk-like, and dense branching morphology-like patterns.
The relation between each pattern and the nutrient and agar concentrations is summarized in the form of a morphological diagram\cite{Matsushita}.
Bacterial colonies have been extensively investigated from the standpoint of physics as a typical pattern-formation phenomenon in which the different patterns appear by changing external conditions.\cite{Vicsek}

According to the measurement of bacterial cells in homogeneously spreading disk-like colonies (formed in soft agar and rich nutrient),
the cell size at the beginning of the expanding phase follows the lognormal distribution $\LN(\ln2.7, 0.24^2)=\LN(0.99, 0.24^2)$, whose median is $\exp(0.99)=\SI{2.7}{\micro\metre}$.
In addition, the cell length in the lag phase has been reported to increase linearly with time up to cell division.
The cell cycle is represented by the lognormal distribution $\LN(\ln22, 0.20^2)=\LN(3.1, 0.20^2)$ (the median is \SI{22}{\minute}), and the growth rate also exhibits lognormal behavior with $\LN(\ln0.078, 0.30^2)=\LN(-2.6, 0.30^2)$ (the median is \SI{0.078}{\micro\metre/\minute}).

\begin{figure}[t!]\centering
\includegraphics[width=8cm]{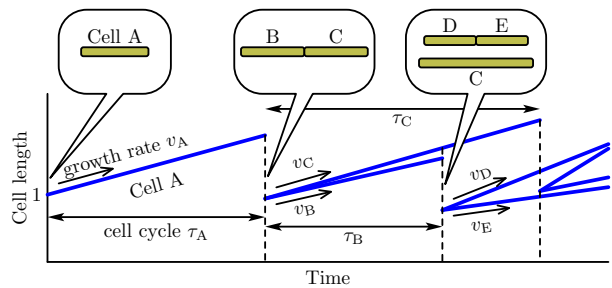}
\caption{(Color online)
Illustration of the growth model for bacterial cells.
An initial cell $\mathrm{A}$ of unit length grows linearly with growth rate $v_\mathrm{A}$, and it divides equally into two cells $\mathrm{B}$ and $\mathrm{C}$ at $t=\tau_\mathrm{A}$.
Cells $\mathrm{B}$ and $\mathrm{C}$ respectively grow linearly with rates $v_\mathrm{B}$ and $v_\mathrm{C}$ until the next cell division, and so forth.
The arrangements of the cells at time $t=0$ (initial state), $t=\tau_\mathrm{A}$ (immediately after the first cell division), and $t=\tau_\mathrm{A}+\tau_\mathrm{B}$ (immediately after the second division) are schematically shown in the balloons.
}
\label{fig1}
\end{figure}

On the basis of these experimental results, a phenomenological model has been proposed.
The procedure of this model is schematically shown in Fig.~\ref{fig1}.
The model starts with a single cell $\mathrm{A}$ of unit length, and assigns cell cycle $\tau_\mathrm{A}$ and growth rate $v_\mathrm{A}$ drawn from lognormal distributions.
Cell $\mathrm{A}$ grows linearly with time, i.e., the size of cell $\mathrm{A}$ at time $t<\tau_\mathrm{A}$ is $1+v_\mathrm{A}t$.
At $t=\tau_\mathrm{A}$, cell $\mathrm{A}$ divides equally into two cells $\mathrm{B}$ and $\mathrm{C}$.
Growth rates $v_\mathrm{B}$ and $v_\mathrm{C}$ and cell cycles $\tau_\mathrm{B}$ and $\tau_\mathrm{C}$ are newly assigned.
For simplicity, the growth rate and cell cycle of each cell are assumed to be independently distributed.
The cells continue their linear growth and cell division in the same way.
According to a numerical calculation\cite{Wakita}, the cell-size distribution becomes stationary after a long time.
By choosing realistic parameters in the lognormal distributions for the growth rate and cell cycle [namely, $\LN(3.1, 0.20^2)$ for the cell cycle and $\LN(-2.6, 0.30^2)$ for the growth rate],
the resultant stationary distribution is described by $\LN(0.92, 0.28^2)$.
This is in good agreement with the actual cell-size distribution $\LN(0.99, 0.24^2)$ of \textit{B.~subtilis}.
Furthermore, it has been reported that the longnormal behavior of the numerical cell-size distribution disappears if the lognormal parameter $\sigma$ of the growth rate or cell cycle is too small or too large.

We theoretically analyze this phenomenological model in the present paper.
We derive the cell size at the onset of the cell cycle in Sect.~\ref{sect2}, and the stationary cell size of the model in Sect.~\ref{sect3}.
They are expressed by sums of lognormal variables, and the corresponding cell-size distributions are not exactly lognormal.
We propose a quantity that evaluates how far the cell size is different from the lognormal behavior.
We numerically show that the deviation from lognormality depends on the product of the growth rate and cell cycle, and derive its scaling property.
From these results, the cell-size distribution of \textit{B.~subtilis} is suggested to be only a lognormal-like distribution, but it can be approximated well by a lognormal distribution owing to the parameter values of the growth rate and cell cycle, which give a good lognormal approximation.

\section{Analysis 1: Cell Size Immediately after Cell Division}\label{sect2}
The aim of this study is to obtain the stationary cell-size distribution of the model, but it is complicated to derive it without preparation.
At each moment, there exist cells that have just divided, divided some earlier, and are about to divide.
We need to consider the variation of the time elapsed from the previous division.
Before studying this cell-size distribution, we start with the cell-size distribution limited to the cells that have just divided, which is a simpler problem.

We focus on the cell size at the onset of the cell cycle and introduce a random variable $X'_n$ as the initial cell size at the $n$th cell cycle.
By setting $V_n$ as the $n$th growth rate and $T_n$ as the $n$th cell cycle, the growth increment during the $n$th cycle is expressed as $V_nT_n$.
The cell size at the end of this cycle is $X'_n+V_nT_n$.
This length is split in half at the cell division, so $X'_{n+1}$ is given by
\begin{equation}
X'_{n+1}=\frac{1}{2}X'_n+\frac{1}{2}V_nT_n.
\label{eq2}
\end{equation}
We easily obtain its solution as
\begin{equation}
X'_n=\frac{1}{2^{n-1}}+\sum_{k=1}^{n-1}\frac{V_kT_k}{2^{n-k}}
\label{eq3}
\end{equation}
by applying Eq.~\eqref{eq2} recursively and using the initial condition $X'_1=1$.
Note that the effect of the initial length $X'_1$ vanishes exponentially as $n$ increases.

The random variables $V_k$ and $T_k$ always appear in the form of the product $V_kT_k$ throughout this paper, and we set a new random variable $L_k:=V_kT_k$.
In light of the experiment, $V_k$ and $T_k$ in the model are lognormal variables.
The product $L_k$ is therefore a lognormal variable, because the product of two independent lognormal variables again follows a lognormal distribution~\cite{Crow}.
We set $\LN(\mu_L, \sigma_L^2)$ for the distribution of $L_k$ and investigate the properties of $X'_n$ in terms of $\mu_L$ and $\sigma_L$.
The model assumes that $\{V_k\}$ and $\{T_k\}$ are independent, and hence $\{L_k\}$ are independently and identically distributed.
Hence, $X'_n$ is given by the sum of independent lognormal variables.

\begin{figure}[t!]\centering
\includegraphics[clip]{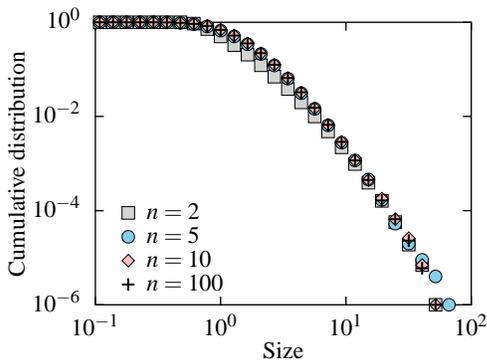}
\caption{(Color online)
Cumulative distributions of $X'_n$ for $n=2$ (squares), 5 (circles), 10 (diamonds), and 100 (crosses).
Each distribution is made up of $10^6$ independent samples, where $\mu_L=0$ and $\sigma_L=1$.
}
\label{fig2}
\end{figure}

By using the mean and variance
\begin{subequations}
\begin{align}
&E[L_1]=E[L_2]=\cdots=e^{\mu_L+\sigma_L^2/2}, \label{eq4a}\\
&V[L_1]=V[L_2]=\cdots=e^{2\mu_L+\sigma_L^2}(e^{\sigma_L^2}-1), \label{eq4b}
\end{align}
\end{subequations}
of $L_k$, the mean and variance of $X'_n$ are respectively calculated as
\begin{align*}
E[X'_n] &= \frac{1}{2^{n-1}} + \sum_{k=1}^{n-1}\frac{E[L_k]}{2^{n-k}}\\
&= e^{\mu_L+\sigma_L^2/2} - \frac{e^{\mu_L+\sigma_L^2/2}-1}{2^{n-1}},\\
V[X'_n] &= \sum_{k=1}^{n-1}\frac{V[L_k]}{4^{n-k}}\\
&= \frac{e^{2\mu_L+\sigma_L^2}(e^{\sigma_L^2}-1)}{3}-\frac{e^{2\mu_L+\sigma_L^2}(e^{\sigma_L^2}-1)}{3\cdot 4^{n-1}}.
\end{align*}
They converge exponentially as $n\to\infty$.
As shown in Fig.~\ref{fig2}, the distribution of $X'_n$ also rapidly becomes stationary;
the distributions for $n=10$ and $100$ appear to be almost identical.
In the numerical calculations below, therefore, we use the distribution for $n=10$ as a substitute for the stationary distribution.

The variable $X'_n$ is given by the weighted sum of independent and identical lognormal variables $L_k$.
A sum of lognormal variables does not exactly follow a lognormal distribution.
(In contrast, a sum of normal variables again follows a normal distribution.)
However, such a sum of variables can be approximated by a single lognormal distribution.
Among the many techniques to approximate a lognormal sum by a single lognormal distribution~\cite{Schwartz, Ho, Mehta},
the simplest and fastest method is the Fenton-Wilkinson (FW) approximation~\cite{Fenton},
in which the lognormal parameters are estimated by matching the first and second moments.
When sample values $\{x_1,\dots,x_N\}$ of $X'_n$ are given, the best lognormal distribution $\LN(\hat\mu, \hat\sigma^2)$ in the FW sense is determined by
\begin{align*}
e^{\hat\mu+\hat\sigma^2/2} &= \frac{x_1+\cdots+x_N}{N} =: m_1,\\
e^{2(\hat\mu+\hat\sigma^2)} &= \frac{x_1^2+\cdots+x_N^2}{N} =: m_2.
\end{align*}
The left-hand sides are the first and second moments of $\LN(\hat\mu, \hat\sigma^2)$.
The estimates $\hat\mu$ and $\hat\sigma$ are explicitly written as
\begin{equation}
\hat\mu = 2\ln m_1 -\frac{1}{2}\ln m_2, \quad
\hat\sigma^2 = \ln m_2 - 2\ln m_1.
\label{eq5}
\end{equation}
In addition to its simplicity, the FW method is known to closely approximate the tail of the cumulative distribution\cite{AbuDayya}.
Since we mainly focus on the cumulative distribution, the FW method is suitable for this study.

We numerically investigated whether $X'_n$ is approximated by a lognormal distribution.
For each $\mu_L$ and $\sigma_L$, we generated independent samples $\{x_1,\dots,x_N\}$ of $X'_n$ by using Eq.~\eqref{eq3} and estimated $\hat\mu$ and $\hat\sigma$ by the FW approximation~\eqref{eq5}.
To evaluate the validity of the lognormal approximation,
we calculated the difference between the estimated lognormal distribution $F_{\hat\mu, \hat\sigma}$ and the empirical cumulative distribution of $\{x_1,\dots,x_N\}$ defined by
\[
G(x) = \frac{\text{number of elements $\ge x$}}{N}.
\]
We define the difference as
\[
D=\left[\int_0^\infty (F_{\hat\mu,\hat\sigma}(x)-G(x))^2 dx\right]^{1/2}.
\]
When the set of samples $\{x_1,\dots,x_N\}$ closely fits the lognormal distribution, $D$ becomes small.
We use cumulative distributions $F_{\hat\mu, \hat\sigma}$ and $G$ in the definition of $D$, because we are primarily interested in whether the tail of $G(x)$ exhibits lognormal decay.
For simplicity, we assume that $\{x_1,\dots,x_N\}$ is arranged in descending order ($x_1\ge\cdots\ge x_N$), which yields $G(x_i)=i/N$.
We replace the integral of $D$ with the Riemann sum
\begin{align*}
\Delta &= \left[\sum_{i=1}^{N-1} \left(F_{\hat\mu,\hat\sigma}(x_i)-G(x_i)\right)^2(x_i-x_{i+1})\right]^{1/2}\\
&=\left[\sum_{i=1}^{N-1} \left(F_{\hat\mu,\hat\sigma}(x_i)-\frac{i}{N}\right)^2(x_i-x_{i+1})\right]^{1/2}.
\end{align*}
This is a discrete form of $D$.

\begin{figure}[t!]\centering
\raisebox{4.5cm}{(a)}
\includegraphics[clip]{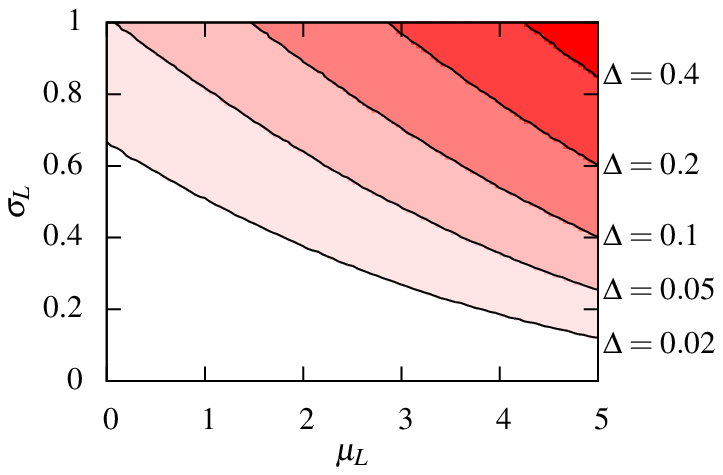}\\
\vspace{5mm}
\raisebox{4.5cm}{(b)}
\includegraphics[clip]{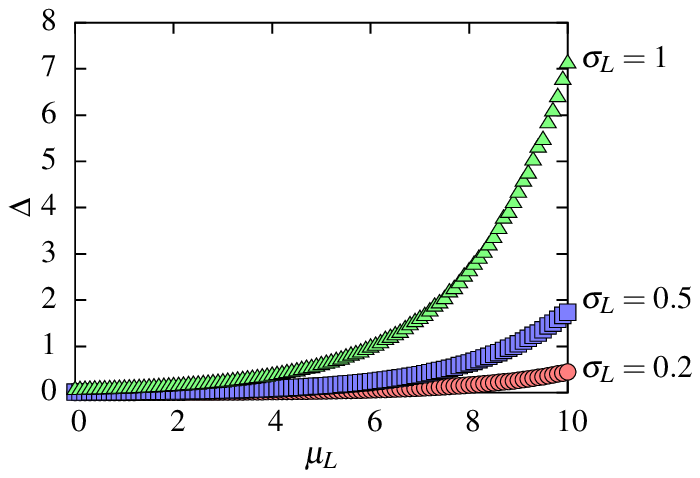}\\
\vspace{5mm}
\raisebox{4.5cm}{(c)}
\includegraphics[clip]{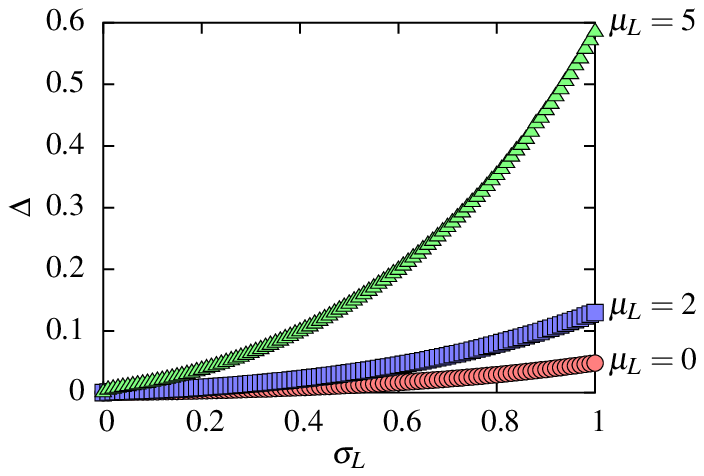}
\caption{(Color online)
Quality of the FW approximation to the distribution of $X'_n$ ($n=10$).
(a) Contour plot of $\Delta$ in the $(\mu_L, \sigma_L)$ plane.
Contours of $\Delta = 0.02, 0.05, 0.1, 0.2$, and $0.4$ are shown by the solid curves.
(b) Graphs of $\Delta$ vs $\mu_L$ at $\sigma_L=0.2$, 0.5, and 1.
(c) Graphs of $\Delta$ vs $\sigma_L$ at $\mu_L=0$, 2, and 5. 
}
\label{fig3}
\end{figure}

In Fig.~\ref{fig3}, we illustrate the numerical result of how close the stationary distribution of $X'_n$ is to the lognormal distribution.
We generated $10^4$ independent samples of $X'_{10}$ using Eq.~\eqref{eq3} (recall that the distribution of $X'_n$ appears to become stationary even at $n=10$), estimated $\hat\mu$ and $\hat\sigma$ by the FW method, and computed $\Delta$.
Figure~\ref{fig3}(a) is a contour plot of $\Delta$ in the $(\mu_L, \sigma_L)$ plane.
The distribution of $X'_{10}$ deviates from the lognormal distribution (i.e., $\Delta$ becomes large) as $\sigma_L$ and $\mu_L$ increase.
Figures~\ref{fig3}(b) and \ref{fig3}(c) show profile curves at fixed $\sigma_L$ ($\sigma_L= 0.2, 0.5$, and $1$) and $\mu_L$ ($\mu_L=0, 2$, and $5$), respectively.

\begin{figure}[t]\centering
\raisebox{4.5cm}{(a)}
\includegraphics[clip]{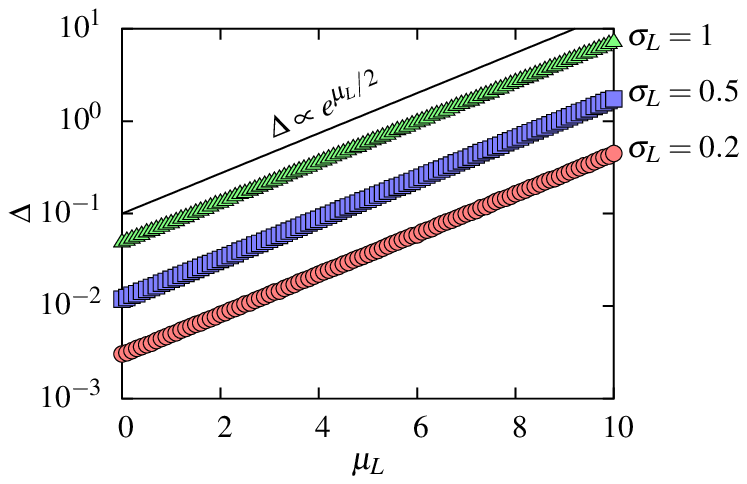}\\
\vspace{5mm}
\raisebox{4.5cm}{(b)}
\includegraphics[clip]{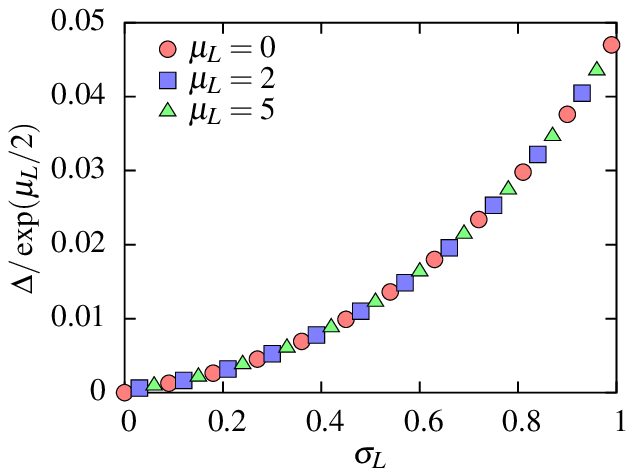}
\caption{(Color online)
Scaling property of $\Delta$.
(a) $\Delta$ grows exponentially with $\mu_L$. The relation $\Delta\propto\exp(\mu_L/2)$ is shown by the solid line.
(b) The graphs in Fig.~\ref{fig3}(b) collapse to a single curve by scaling $\Delta$ by the factor $\exp(\mu_L/2)$.
Different points are alternately aligned to avoid overlapping.
}
\label{fig4}
\end{figure}

The dependence of $\Delta$ on $\mu_L$ [Fig.~\ref{fig3}(b)] can be explained briefly as follows.
If we multiply $\{x_1,\dots,x_N\}$ by a factor $e^\alpha$ to obtain $\{e^\alpha x_1,\dots,e^\alpha x_N\}$,
the lognormal estimates become $\hat\mu+\alpha$ and $\hat\sigma^2$.
Since the scaling relation $F_{\hat\mu+\alpha,\hat\sigma}(e^\alpha x_i)=F_{\hat\mu,\hat\sigma}(x_i)$ holds from Eq.~\eqref{eq1},
$\Delta$ for the sample $\{e^\alpha x_i\}$ is expressed as
\begin{align*}
&\Delta(\{e^\alpha x_i\})\\
&=\left[\sum_{i=1}^N\left(F_{\hat\mu+\alpha,\hat\sigma}(e^\alpha x_i)-\frac{i}{N}\right)^2(e^\alpha x_{i+1}-e^\alpha x_i)\right]^{1/2}\\
&=\left[e^\alpha\sum_{i=1}^N\left(F_{\hat\mu,\hat\sigma}(x_i)-\frac{i}{N}\right)^2(x_{i+1}-x_i)\right]^{1/2}\\
&=e^{\alpha/2}\Delta(\{x_i\}).
\end{align*}
Hence, multiplying $\{x_i\}$ by $e^\alpha$ causes the factor $\exp(\alpha/2)$ to $\Delta$.
Meanwhile, multiplying $\{x_i\}$ by $e^\alpha$ corresponds to replacing $L_k(=V_kT_k)$ in Eq.~\eqref{eq3} with $e^\alpha L_k$.
Since $e^\alpha L_k$ follows $\LN(\mu_L+\alpha,\sigma_L^2)$, using $e^{\alpha}L_k$ instead of $L_k$ is equivalent to adding $\alpha$ to $\mu_L$.
Finally, $\Delta\propto\exp(\mu_L/2)$ is derived.
We numerically confirmed this relation (see Fig.~\ref{fig4}).
Figure~\ref{fig4}(a) shows that the graphs in Fig.~\ref{fig3}(b) grow exponentially.
Thus, the three graphs in Fig.~\ref{fig3}(c) collapse into a single curve by using the scaled value $\Delta/\exp(\mu_L/2)$, as shown in Fig.~\ref{fig4}(b).
In short, the increase in $\Delta$ against $\mu_L$ is simply due to the scaling effect of $\Delta$.

\section{Analysis 2: Stationary Cell-Size Distribution in the Model}\label{sect3}
In this section, we investigate the cell-size distribution in the model after a long time, which also corresponds to the experimental cell-size distribution.
Note that this distribution involves all cells, not only cells just after division.
Thus, the distribution is different from $X'_n$ stated in the previous section.

After a long time, during which the cells undergo division many times, the cell size just after the division is written as
\begin{equation}
X'_\infty = \sum_{k=1}^\infty \frac{L_k}{2^k}.
\label{eq6}
\end{equation}
This expression is obtained by taking the limit $n\to\infty$ in Eq.~\eqref{eq3}.
In this stationary state, the cell size just before the division is given by $L_0+X'_\infty$, where $L_0$ is a lognormal variable with $\LN(\mu_L, \sigma_L^2)$.
At an arbitrary time, a randomly chosen cell takes a uniformly random size between $X'_\infty$ and $L_0+X'_\infty$.
The size of a cell randomly chosen at an arbitrary time is therefore written as
\begin{equation}
X = rL_0 + \sum_{k=1}^\infty \frac{L_k}{2^k},
\label{eq7}
\end{equation}
where $r$ is a uniform random number on the unit interval $[0,1]$ and $L_0, L_1,\ldots$ are independently and identically distributed with the lognormal distribution $\LN(\mu_L, \sigma_L^2)$.
The first term $rL_0$ on the right-hand side represents the fluctuation of the time elapsed from the previous cell division.
Note that $r=0$ and $r=1$ respectively correspond to the cell sizes just after division and just before division.

\begin{figure}[t]\centering
\includegraphics[clip]{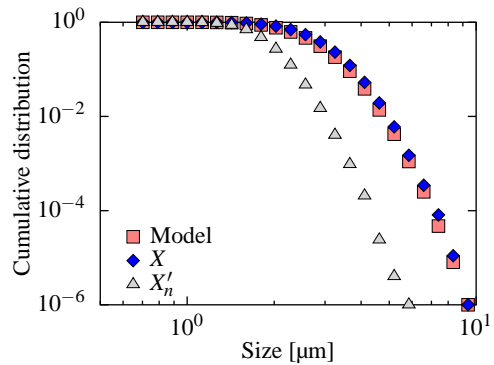}
\caption{(Color online)
Cell-size distributions obtained by direct simulation using the model (squares), and of $X$ (diamonds) and $X'_\infty$ (triangles).
The distribution of $X$ is close to that obtained by direct simulation, but the distribution of $X'_\infty$ is not.
}
\label{fig5}
\end{figure}

We numerically verified that Eq.~\eqref{eq7} gives a cell-size distribution corresponding to the model (see Fig.~\ref{fig5}).
The square plots show the cell-size distribution obtained by directly simulating the time evolution of cells along the procedure of the model (as in Fig.~\ref{fig1}).
The cell cycles and growth rates were drawn from the lognormal distributions $\LN(3.1, 0.20^2)$ and $\LN(-2.6, 0.30^2)$, respectively.
We computed the cell size distribution when the total number of cells became $10^6$.
The diamonds show the distribution of $X$ obtained from $10^6$ samples, in which we replaced the infinite sum in Eq.~\eqref{eq7} with the first ten terms.
These two graphs are very close to each other, which indicates that Eq.~\eqref{eq7} is a valid expression.
The distribution of $X'_n$ ($n=10$) is shown as the triangles, but is obviously different from the other distributions.

\begin{figure*}\centering
\raisebox{4.5cm}{(a)}
\includegraphics[clip]{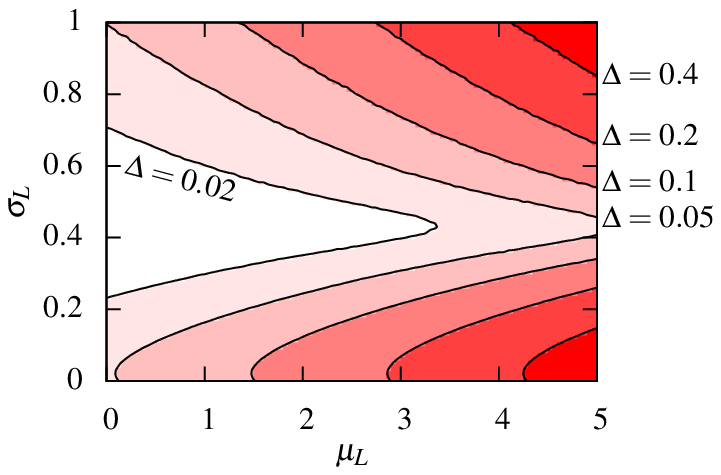}
\hspace{5mm}
\raisebox{4.5cm}{(b)}
\includegraphics[clip]{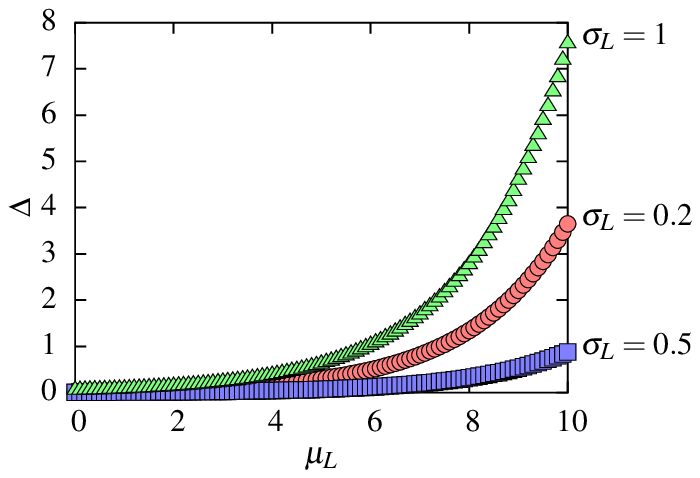}\\
\vspace{5mm}
\raisebox{4.5cm}{(c)}
\includegraphics[clip]{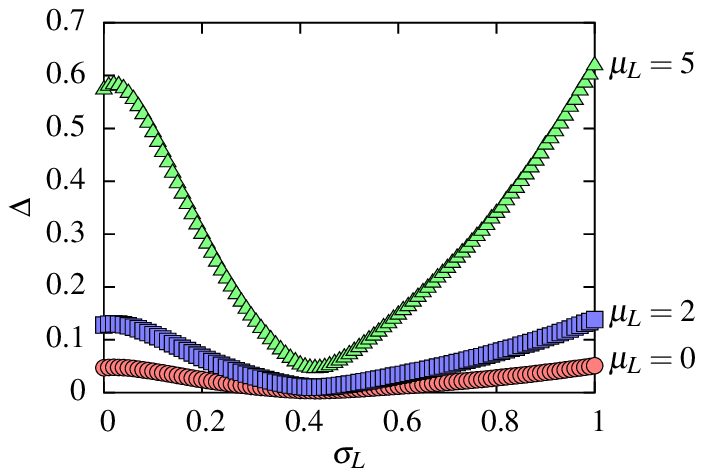}
\hspace{5mm}
\raisebox{4.5cm}{(d)}
\includegraphics[clip]{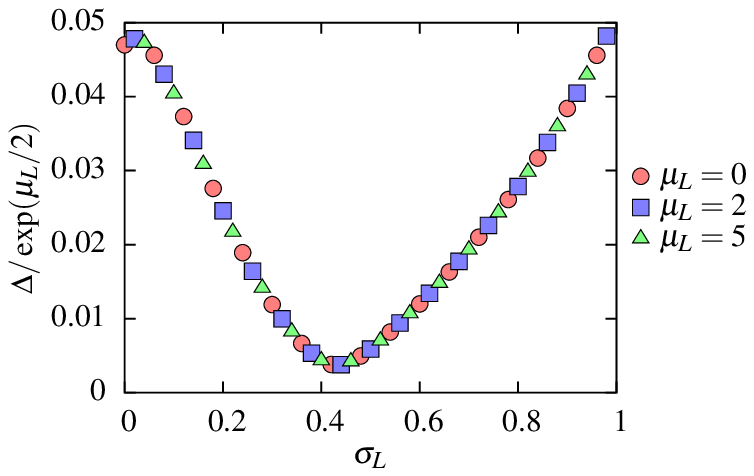}
\caption{(Color online)
Properties of $\Delta$ for the distribution of $X$.
(a) Contour plot of $\Delta$ in the $(\mu_L, \sigma_L)$ plane.
(b), (c) Profile curves at fixed $\sigma_L$ ($\sigma_L = 0.2, 0.5,$ and 1) and $\mu_L$ ($\mu_L = 0, 2,$ and 5), respectively.
(d) The curves in (c) overlap each other upon using the scaled value $\Delta/\exp(\mu_L/2)$.
As with Fig.~\ref{fig4}(b), different points are aligned alternately.
}
\label{fig6}
\end{figure*}

It is not clear whether the distribution of $X$ is approximated by a single lognormal distribution.
We show in Fig.~\ref{fig6} the validity of the lognormal approximation of $X$ by the FW method.
As with Fig.~\ref{fig3}, we calculated $\Delta$ by using $10^4$ samples of $X$, in which the infinite sum in Eq.~\eqref{eq7} was replaced with the first ten terms.
Figure~\ref{fig6}(a) is a contour plot of $\Delta$ in the $(\mu_L, \sigma_L)$ plane.
The shape of the contour lines is more intricate than that in Fig.~\ref{fig3}(a).
Figures~\ref{fig6}(b) and \ref{fig6}(c) respectively illustrate profile curves at fixed $\sigma_L$ ($\sigma_L = 0.2, 0.5,$ and 1) and $\mu_L$ ($\mu_L = 0, 2,$ and 5).
The curves in Fig.~\ref{fig6}(c) are not monotonic and they all take the minimum values at $\sigma_L=0.43$--0.44.
That is, the distribution of $X$ is greatly inconsistent with a lognormal distribution when $\sigma_L$ is too small or too large.
Upon using the scaled value $\Delta/\exp(\mu_L/2)$, the curves in Fig.~\ref{fig6}(c) collapse into a single curve [see Fig.~\ref{fig6}(d)].

\begin{figure}[t]\centering
\includegraphics[clip]{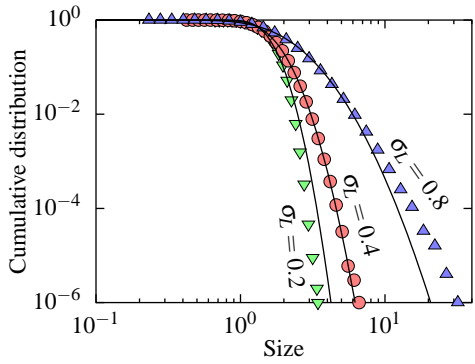}
\caption{(Color online)
Comparison of distributions of $X$ with $\mu_L=0$ and $\sigma_L=0.2$ (lower triangles), 0.4 (circles), and 0.8 (upper triangles).
The solid curves are the lognormal approximation results obtained by the FW method.
}
\label{fig7}
\end{figure}

In Fig.~\ref{fig7}, the distributions of $X$ with $\mu_L=0$ and $\sigma_L=0.2$ (lower triangles), 0.4 (circles), and 0.8 (upper triangles) are presented.
The lognormal distributions obtained by the FW approximation are shown by the solid curves.
The distribution of $\sigma_L=0.4$ [near the bottom of the curve in Fig.~\ref{fig6}(d)] perfectly fits the lognormal distribution.
In contrast, the distribution of $\sigma_L=0.2$ decays faster than the lognormal distribution, and that of $\sigma_L=0.8$ decays slower than the lognormal.

In the actual growth of \textit{B.~subtilis},
the growth rate $V_k$ and cell cycle $T_k$ respectively obey $\LN(\ln22, 0.20^2)$ and $\LN(\ln0.078, 0.30^2)$.
The distribution of their product $L_k=V_kT_k$ is then given by
\[
\LN(\ln22+\ln0.078, 0.20^2+0.30^2)=\LN(0.54, 0.36^2).
\]
Note that $\sigma_L=0.36$ is near the bottom of the curve in Fig.~\ref{fig6}(d), at which $\Delta$ becomes small.
This is a reason why the actual cell-size distribution of \textit{B.~subtilis} agrees very well with a lognormal distribution.

The mean of $X$ is easily calculated as
\[
E[X]=E[r]E[L_0]+\sum_{k=1}^\infty\frac{E[L_k]}{2^k}=\frac{3}{2}e^{\mu_L+\sigma_L^2/2},
\]
where we used Eq.~\eqref{eq4a} and $E[r]=1/2$ (mean of the uniform random number on $[0,1]$).
The variance of $X$ is calculated as
\[
V[X]=V[rL_0]+\sum_{k=1}^\infty\frac{V[L_k]}{4^k}
= \frac{2}{3}e^{2\mu_L+2\sigma_L^2} - \frac{7}{12}e^{2\mu_L+\sigma_L^2},
\]
where we used Eq.~\eqref{eq4b} and the formula\cite{Goodman}
\[
V[rL_0] = V[r]V[L_0] + E[r]^2V[L_0] + V[r]E[r]^2
\]
for the variance of the product, with $V[r]=1/12$.
The square mean of $X$ is then given by
\[
E[X^2] = V[X] + E[X]^2 = \frac{2}{3}e^{2\mu_L+2\sigma_L^2}+\frac{5}{3}e^{2\mu_L+\sigma_L^2}.
\]
The parameter $\hat\mu$ of the FW approximation is
\[
\hat\mu = 2\ln E[X] - \frac{1}{2}\ln E[X^2],
\]
where we substitute $E[X]$ and $E[X^2]$ for $m_1$ and $m_2$ in Eq.~\eqref{eq5}, respectively.
We then obtain the median of $X$ as
\[
e^{\hat\mu} = \frac{E[X]^2}{E[X^2]^{1/2}}
= \frac{9\sqrt{3}}{4}\frac{e^{\mu_L}}{(2+5e^{-\sigma_L^2})^{1/2}}.
\]
With this result, we can compute the median of the stationary cell size $X$ from two parameters, $\mu_L$ and $\sigma_L$.
By using the values $\mu_L=0.54$ and $\sigma_L=0.36$ from the experiment, we have
\[
e^{\hat\mu} \approx \SI{2.65}{\micro\metre}.
\]
This estimate correctly gives the experimental median cell size of \SI{2.7}{\micro\metre}.

\section{Discussion}
In this paper, we have solved the phenomenological stochastic model for bacterial growth and have shown that the stationary cell size \eqref{eq7} is expressed by using lognormal variables and a uniform variable.
This result suggests that the cell size of \textit{B.~subtilis} does not follow a genuine lognormal distribution but a lognormal-like distribution.
The parameter $\sigma_L$ is an important indicator of the quality of the lognormal approximation.
This study has elucidated that the value $\sigma_L=0.36$ for actual bacterial growth, which is near the bottom of Fig.~\ref{fig6}(d) ($\sigma_L\approx0.43$), is the reason why the experimental cell-size distribution can be approximated well by a lognormal distribution.
The value of $\sigma_L$ can be varied by changing external conditions
 or by using other bacterial species such as \textit{Escherichia~coli}\cite{Fujihara} and \textit{Proteus~mirabilis}\cite{Matsuyama}.
We consider that the model and analysis in this study should be quantitatively inspected by estimating $\sigma_L$ and $\Delta$ from experiments under various conditions.

\begin{figure}[t!]\centering
\includegraphics[clip]{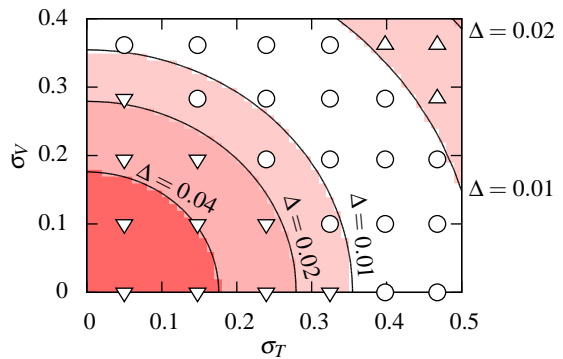}
\caption{(Color online)
Quality of the lognormality of the stationary cell-size distribution in the $(\sigma_T, \sigma_V)$ plane.
The circles (showing that the lognormal approximation is valid) and upper and lower triangles (lognormal approximation is invalid) are results of Wakita \textit{et al.}\cite{Wakita}.
Contours of $\Delta=0.01, 0.02$, and 0.04 are shown by solid curves.
}
\label{fig8}
\end{figure}

We make a further comparison between our result and a numerical result of Wakita \textit{et al.}\cite{Wakita} in which the cell cycle $T_k$ and growth rate $V_k$ are varied separately.
Here we set $\LN(\mu_T, \sigma_T^2)$ and $\LN(\mu_V, \sigma_V^2)$ for the distributions of $T_k$ and $V_k$, respectively.
Wakita \textit{et al.}\cite{Wakita} investigated the lognormality of the stationary cell-size distribution of the model in the $(\sigma_T, \sigma_V)$ plane, instead of $\sigma_L$, with fixed medians $\exp(\mu_T)=\SI{20}{\minute}$ and $\exp(\mu_V)=\SI{0.1}{\micro\metre/\minute}$.
This result is shown by the circles (the lognormal approximation is valid) and upper and lower triangles (the lognormal approximation is not good) in Fig.~\ref{fig8}.
The size distribution deviates from the lognormal distribution for small $\sigma_T$ and $\sigma_V$ or for large $\sigma_T$ and $\sigma_V$.
In this figure, the contours of $\Delta = 0.01, 0.02$, and 0.04 are also shown by the solid curves.
The relation $\sigma_L^2=\sigma_T^2+\sigma_V^2$ is obtained from $L_k=V_kT_k$,
and hence, the curve of $\sigma_L=\text{const.}$ forms an arc centered at the origin in the $(\sigma_T, \sigma_V)$ plane.
The value of $\Delta$ depends on $\sigma_L$ and not on $\sigma_T$ and $\sigma_V$;
thus, the contours of $\Delta$ are concentric circles.
In contrast, the boundary between the circles and the lower triangles seems to be a straight line.
We deduce that this difference arises because the judgment of a circle or triangle relied on human eyes.
Nevertheless, we consider our result to be consistent with the previous study as a whole.
In fact, the region of $\Delta<0.01$ contains only circles, and that of $\Delta>0.02$ contains only lower triangles.
Moreover, circles and lower triangles both exist in the narrow range $0.01<\Delta<0.02$.

The model assumes the lognormality of the growth rate $V_k$ and cell cycle $T_k$.
(Theoretically, the lognormality of their product $L_k=V_kT_k$ is essential in our analysis.)
As a matter of fact, however, it is unclear why they follow lognormal distributions.
At the same time, we cannot exclude the possibility that they actually follow lognormal-like distributions.
The stochastic properties of $V_k$ and $T_k$ may involve dynamics inside a cell, and this is a problem for future research.
Increasing the amount of experimental data can also help to investigate $V_k$ and $T_k$ in more detail.

The difference $\Delta$ between a cell-size distribution and its FW approximate becomes large when $\sigma_L$ increases in both Figs.~\ref{fig3}(a) and \ref{fig6}(a).
This behavior has also been found for a finite sum of lognormal variables\cite{Schwartz}.
For small $\sigma_L$, on the other hand, $\Delta$ for $X$ [Fig.~\ref{fig6}(a)] is greatly different from that for $X'_\infty$ [Fig.~\ref{fig3}(a)].
Let us consider the limiting case $\sigma_L=0$,
that is, the random variable $L_k$ is constant [$L_k\equiv\exp(\mu_L)$].
$X'_\infty$ in Eq.~\eqref{eq6} is also constant [$X'_\infty\equiv\exp(\mu_L)$]
and $X'_\infty$ is distributed lognormally with $\LN(\mu_L, 0)$.
Thus, $\Delta=0$ for $X'_\infty$.
[We can observe $\Delta\to0$ as $\sigma_L\to0$ in Fig.~\ref{fig3}(c).]
On the other hand, the random variable $rL_0$ in Eq.~\eqref{eq7} is uniformly distributed in the interval $[0, \exp(\mu_L)]$, and $X(=rL_0+X'_\infty)$ is uniformly distributed on $[\exp(\mu_L), 2\exp(\mu_L)]$.
The uniform distribution is different from a lognormal distribution, so $\Delta$ for $X$ does not become zero as $\sigma\to0$.
Thus, the property of $\Delta$ for small $\sigma_L$ is affected by the term $rL_0$ in Eq.~\eqref{eq7}.
We conjecture that the balance between the increase in $\Delta$ for large $\sigma_L$ and the existence of the term $rL_0$ corresponds to the minimum point in Fig.~\ref{fig6}(d), 
but we have not yet carried out a detailed analysis, especially of why the minimum is at $\sigma_L\approx0.43$.

In order to comprehend complex systems, we usually describe a phenomenon broadly at first, neglecting its details, then gradually increase the accuracy.
In this paper, we have investigated a measure of how much an actual distribution deviates from the approximate lognormal distribution, which is an in-depth expression of the result that the cell-size distribution is close to the lognormal distribution.
Lognormal behavior has been reported in many systems as a first approximation, and we hope that our analysis will help promote further research on such systems.

\begin{acknowledgments}
The authors are very grateful to H. R. Brand for the fruitful discussion of the manuscript.
This work was supported by a Grant-in-Aid for Young Scientists (B) (25870743) from the Japan Society for the Promotion of Science (KY), and a Chuo University Grant for Special Research and a Grant-in-Aid for Exploratory Research (15K13537) from JSPS (JW).
\end{acknowledgments}


\begin{thebibliography}{33}
\bibitem{Newman}
M. E. J. Newman, Contemp. Phys. 46, 323 (2005).
\bibitem{Buchanan}
M. Buchanan, Ubiquity: Why Catastrophes Happen (Three Rivers Press, New York, 2000).
\bibitem{Crow}
E. L. Crow and K. Shimizu, Lognormal Distributions (Marcel Dekker, New York, 1988).
\bibitem{Kobayashi}
N. Kobayashi, H. Kuninaka, J. Wakita, and M. Matsushita, J. Phys. Soc. Jpn. 80, 072001 (2011).
\bibitem{Limpert}
E. Limpert, W. A. Stahel, and M. Abbt, BioScience 51, 341 (2001).
\bibitem{Katsuragi}
H. Katsuragi, D. Sugino, and H. Honjo, Phys. Rev. E 70, 065103(R) (2004).
\bibitem{Bittelli}
M. Bittelli, G. S. Campbell, and M. Flury, Soil Sci. Soc. Am. J. 63, 782 (1998).
\bibitem{Uttley}
P. Uttley, I. M. McHardy, and S. Vaughan, Mon. Not. R. Astron. Soc. 359, 345 (2005).
\bibitem{Furusawa}
C. Furusawa, T. Suzuki, A. Kashiwagi, T. Yomo, and K. Kaneko, Biophysics 1, 25 (2005).
\bibitem{Sasaki}
Y. Sasaki, H. Kuninaka, and M. Matsushita, J. Phys. Soc. Jpn. 76, 074801 (2007).
\bibitem{Redner}
S. Redner, Phys. Today 58, 49 (2005).
\bibitem{Fortunato}
S. Fortunato and C. Castellano, Phys. Rev. Lett. 99, 138701 (2007).
\bibitem{Abramowitz}
M. Abramowitz and I. A. Stegun, Handbook of Mathematical Functions (Dover, New York, 1965).
\bibitem{Sornette}
D. Sornette, Critical Phenomena in Natural Sciences (Springer, Berlin, 2006).
\bibitem{Mitzenmacher}
M. Mitzenmacher, Internet Math. 1, 226 (2004), 2nd ed.
\bibitem{Malevergne}
Y. Malevergne, V. Pisarenko, and D. Sornette, Phys. Rev. E 83, 036111 (2011).
\bibitem{Kuninaka}
H. Kuninaka, Y. Mitsuhashi, and M. Matsushita, J. Phys. Soc. Jpn. 78, 125001 (2009).
\bibitem{Stumpf}
M. P. H. Stumpf and P. J. Ingram, Europhys. Lett. 71, 152 (2005).
\bibitem{Takayasu}
H. Takayasu, A. Sato, and M. Takayasu, Phys. Rev. Lett. 79, 966 (1999).
\bibitem{Manrubia}
S. C. Manrubia and D. H. Zanette, Phys. Rev. E 59, 4945 (1999).
\bibitem{Yamamoto2012}
K. Yamamoto and Y. Yamazaki, Phys. Rev. E 85, 011145 (2012).
\bibitem{Yamamoto2014}
K. Yamamoto, Phys. Rev. E 89, 042115 (2014).
\bibitem{Wakita}
J. Wakita, H. Kuninaka, T. Matsuyama, and M. Matsushita, J. Phys. Soc. Jpn. 79, 094002 (2010).
\bibitem{Matsushita}
M. Matsushita, F. Hiramatsu, N. Kobayashi, T. Ozawa, Y. Yamazaki, and T. Matsuyama, Biofilms 1, 305 (2004).
\bibitem{Vicsek}
T. Vicsek, Fluctuations and Scaling in Biology (Oxford University Press, New York, 2001).
\bibitem{Schwartz}
S. C. Schwartz and Y. S. Yeh, Bell Syst. Tech. J. 61, 1441 (1982).
\bibitem{Ho}
C.-L. Ho, IEEE Trans. Veh. Technol. 44, 756 (1995).
\bibitem{Mehta}
N. B. Mehta, J. Wu, A. F. Molisch, and J. Zhang, IEEE Trans. Wireless Commun. 6, 2690 (2007).
\bibitem{Fenton}
L. F. Fenton, IRE Trans. Commun. Syst. 8, 57 (1960).
\bibitem{AbuDayya}
A. A. Abu-Dayya and N. C. Beaulieu, IEEE Trans. Veh. Technol. 43, 163 (1994).
\bibitem{Goodman}
L. A. Goodman, J. Am. Stat. Assoc. 55, 708 (1960).
\bibitem{Fujihara}
M. Fujihara, J. Wakita, D. Kondoh, M. Matsushita, and R. Harasawa, Afr. J. Microbiol. Res. 7, 1780 (2013).
\bibitem{Matsuyama}
T. Matsuyama, Y. Takagi, Y. Nakagawa, H. Itoh, J. Wakita, and M. Matsushita, J. Bacteriol. 182, 385 (2000).
\end{thebibliography}
\end{document}